\documentstyle[amssymb,amsmath, 12pt]{article}

\def\nd{\noindent}

\begin{document}

\def\slash#1{\setbox0=
\hbox{$#1$}#1\hskip-\wd0\hbox to\wd0{\hss\sl/\/\hss}}

\begin{center}{\large\bf
The Origin of Spontaneous Symmetry Breaking in
Theories with Large Extra Dimensions}
\end{center}
\vskip 1truecm
\centerline{\large
G.\ Dvali\footnote{e-mail: gd23@NYU.EDU},\
S.\ Randjbar--Daemi\footnote{e-mail: daemi@ictp.trieste.it},\
and R.\
Tabbash\footnote{e-mail: rula@sissa.it}
}

\bigskip
\bigskip
\centerline{\it
Dept. of Physics, New York University, New York, NY
10003$^{1}$}

\smallskip
\centerline{\it
International Centre for Theoretical Physics, Trieste,
Italy$^{2}$}

\smallskip
\centerline{\it
SISSA--ISAS,
Via Beirut 4, I-34014 Trieste, Italy$^{3}$}

\vskip 1.2truecm

\begin{abstract}
{

\nd We suggest that the electroweak
Higgs particles can be identified with
extra-dimensional components of the gauge fields, which after
compactification on a certain topologically non-trivial
 background become tachyonic and condense.  If the
tachyonic mass is a tree level effect, the natural scale of the
gauge symmetry breaking is set by the inverse radius of the
internal space, which, in case of the electroweak symmetry, must
be around $\sim 1/$TeV. We discuss the possibility of a vanishing
tree level mass for the Higgs. In such a scenario the tachyonic
mass can be induced by quantum loops and can be naturally smaller
than the compactification scale. We give an example in which this
possibility can be realized. Starting from an
Einstein--Yang--Mills theory coupled to fermions in
$10$-dimensions, we are able to reproduce the spectrum of the
Standard Model like chiral fermions and Higgs type scalars in
$4$-dimensions upon compactifying on ${\mathbb{C}}P^1\times
{\mathbb{C}}P^2$. The existence of a monopole solution on
${\mathbb{C}}P^1$ and a self dual $U(1)$ instanton on
${\mathbb{C}}P^2$ are essential in obtaining chiral fermions
as well as tachyonic or massless scalars in $4$-dimensions.
We give a simple rule which helps us to identify the presence
of tachyons on the monopole background on $S^2$.}
\end{abstract}
\vskip 1truecm
\newpage


\section{Introduction}


Perhaps the most mysterious part of the Standard Model
(SM) is the Higgs
sector, which is responsible for the origin of the
electroweak scale
$M_W\sim 100$GeV.
In conventional $4D$ field theories it is hard to
understand how such a
low (compared to $M_{Planck} \sim 10^{19}$GeV) symmetry
breaking scale
triggered by an elementary scalar could be perturbatively
stable. Thus, in
such a framework one has to rely on either low energy
supersymmetry or
technicolor. An alternative understanding of this
enormous hierarchy may
come from the view that an effective cut-off of the
$4D$ field theory (the
fundamental quantum gravity scale $M$) is not far from
$M_W$, due to
existence of large extra dimensions \cite{add}.\\

 The idea of large extra dimensions explains {\it perturbative
stability} of the weak scale versus the Planck scale, by lowering
the cut-off of the theory. Nevertheless, it does not address the
issue of {\it sensitivity} of the Higgs mass to the ultraviolet
cut-off. This is an  important issue from the point of view of
the low energy calculability, and is the central point to be
addressed in the present work. We propose a framework in which
such a sensitivity is absent as a result of a symmetry and the
Higgs mass is finite and has no dependence on the unknown
ultraviolet physics. Note that in our case the sensitivity is
weaker than in softly broken $N=1$ supersymmetry, where the Higgs
mass is log-sensitive to the cut-off. Such an unusual situation
is possible due to the presence of a symmetry that forbids any
local counter term generating a Higgs mass. 
 In our case the symmetry in question is a
higher-dimensional gauge symmetry, which forbids the mass of 
Higgs particles to receive cut off dependent ( and hence local) corrections,
 due to the fact that the Higgs fields are
identified with some of the components of a higher-dimensional
gauge field. Higher-dimensional gauge symmetry is spontaneously
broken by compactification. {An alternative approach was
discussed in \cite{barbieri}, where the Higgs mass is controlled
by a higher-dimensional extended supersymmetry, spontaneously
broken globally by Scherk-Schwarz mechanism.}

 It should be noted that the idea of large extra dimensions,
{\it per se} does not answer why the electroweak
symmetry should be broken at all,\footnote{For some discussions
see \cite{breaking}} nor it says anything about the origin of the
Higgs particles. Also, many essential issues, such as, for
instance, generation of chiral fermions with needed quantum
numbers, are usually attributed to technicalities. The present
paper describes an attempt to provide a unified solution to all the
above listed issues by implementing some of the ideas of an
earlier work\cite{sss1} in the framework  of standard model and
large extra dimensions.\\

 As mentioned above, we suggest that the Higgs particles may be
identified with the
extra components of the higher-dimensional gauge fields. After
compactification of extra dimensions on a monopole background,
via mechanism of \cite{sss1}, some of the extra components of the
gauge fields become tachyonic and spontaneously break the
electroweak symmetry. Their quantum numbers are identical to those
of SM Higgs doublet. Notice that the monopole background is
essential for generating the families of chiral fermions in
four-dimensions, and therefore is doing a double job. If the
tachyonic mass is a tree level effect the natural scale of the
symmetry breaking is $\sim 1/a$, the inverse radius of extra
compact space, since the only source of spontaneous breaking of
the higher-dimensional gauge invariance is the compactification
itself. In  other words, since in the infinite volume limit $a
\rightarrow \infty$ the full higher-dimensional gauge invariance
must be recovered, the weak scale must go as $M_W \sim 1/a$. Thus
in this case the size of extra dimensions to which gauge fields
can propagate should be $a \sim 1/$TeV (as in \cite{antoniadis}).
However, it is important to stress that in order for the theory
not to become infinitely strongly coupled above the
compactification scale, the cut-off $M$ must be lowered as in
\cite{add}, possibly via increase of the volume of some
additional dimensions to which only gravity can spread. This
issue will not be discussed in the present work.\\

The main idea behind our construction is very simple and goes
back to \cite{sss1}. Consider a gauge group $G$ in
$4+N$-dimensions, with the gauge fields $V_A$. After $N$ extra
dimensions are compactified, the extra $V_m;~~m=5,...,4+N$
components of the gauge field appear as scalars \cite{manton},
and the
remaining $V_{\mu}$-components as gauge fields in
four-dimensional effective theory. In some cases it is necessary
that the compactified manifold involves a non-trivial topology
and gauge connection (e.g. such as monopole configuration) in
order to generate chiral fermions \cite{sss2,witten1}. On such a
background $V_m$ components often become tachyonic unless special
conditions are met. Our idea is to precisely use this instability
for the spontaneous breaking of the gauge symmetry in
four-dimensions. We shall argue that this leads to a solution of the
hierarchy problem.\footnote{We would like to emphasize that by the hierarchy problem here we
do not mean the presence of numerically very different masses
like the Planck  mass of $10^{19}$GeV and the Higgs mass of
about a TeV. What we have in mind  is a mechanism which
generates a finite Higgs mass whose value is stable under quantum
corrections. A well known  way to ensure this stability is to
invoke low energy supersymmetry.  In the present paper we are
essentially replacing the low energy  supersymmetry with local
gauge symmetry.}\\

For  obvious phenomenological reasons, ideally  we would like
to have the electroweak symmetry breaking scale smaller than
$1/a$. As we said above, in the simplest realization of our
scheme the scale comes out just around $1/a$. However, we shall
also discuss some ideas how the one-loop hierarchy $M_W < 1/a$
might be generated. More specifically we shall consider the
possibility that the monopole induced tree level masses of the
scalars vanish. Unless the mass of scalars are protected by
supersymmetry, which we are not considering in this paper, the
loop effects will generally induce a non zero mass for these
fields. With an appropriate choice of the fermionic couplings
this mass can be made tachyonic. Of course in general the loop
induced mass corrections are  divergent and they become infinite
in the limit of infinite cut off.  Unless the divergences vanish for
some symmetry reason we need to renormalize the theory. Here we would
like to argue that  considering the scalars as some components of a
Yang-Mills vector potentials evades the
hierarchy problem which the
conventional renormalization in $4$ dimensions generate.  Our argument goes
as follows.\\

There are two distinct energy scales in our problem. The first is the
Planck mass in higher dimensions which we assume is  a few powers of 10
times a TeV. The second one is the inverse of a typical radius of the compact
dimensions. Clearly as
we explore distances shorter than the size of the compact
dimensions the full
gauge symmetry
will be restored.  This domain can still be larger than the
fundamental Planck length of the higher dimensional theory. In this regime
we shall see no Higgs field. All we see will be massless gauge particles.
 In order for this idea to work it is important that we consider our theory
as an effective theory which includes all terms compatible with the gauge
symmetry  and the general coordinate invariance. This will require the
presence of infinite number of parameters in our effective
Lagrangian. Thus
any cut off dependence can be absorbed in the redefinition of these
parameters. The Higgs mass then should be written in the generic form of
$\displaystyle\mu^2 = \frac{1}{a^2} f(Ea)$, where
$a$ is a typical radius in our model and $E$ is some common energy scale.
The main task is to compute the function $f$, which in this paper we
calculate only at the tree level.  Our statement is that since  at
energies much larger than
$1/a$ the full gauge symmetry is recovered, the Higgs mass can not
be heavier than $1/a$.
This is our suggestion to solve the Higgs hierarchy
problem.\\

The function $f$ has been calculated in \cite{h}
for toroidal internal spaces
and have been shown to be finite.
These references contain ideas similar to the one advocated in the present
paper. However, we would like to draw the reader's attention
to a basic difference, namely, the very important and central role
of topologically non-trivial gauge field backgrounds, such as
instantons in our work. In \cite{bachas}, similar
ideas have been used to study dynamical breaking of supersymmetry
in the context of type I string theory. \\

In this paper, we shall construct {\it non}-GUT type models,
i.e. in the effective
$4$-dimensional
theory the
leptons and quarks will
not be in the same multiplet of the
gauge group.
This makes the task of model building considerably
more delicate. We have to find an appropriate gauge field
configuration on a compact space which gives rise to the
intricate chiral structure of one family of quarks and leptons.
Furthermore we need to ensure that the tachyonic Higgs fields are
singlets of the color $SU(3)$ and doublets of $SU(2)_L$. This is
difficult because in the Kaluza--Klein approach all the
4-dimensional fields have the same parents in the higher
dimensions and the mechanism should be subtle enough to produce
the rich spectrum of the $4$-dimensional standard model or its
generalization. Because of the bigger multiplet structure of the
GUT models the problem is somewhat easier if one aims to obtain a
GUT
model. \\

In this paper we shall start from a $6$ or $10$ dimensional
theory and try to obtain models in $D=4$ for which the gauge
group contains the standard model $U(1)\times SU(2)\times SU(3)$
with the correct fermion and Higgs sector.\footnote{ In the
$D=10$ example there will be two different $U(1)_Y$, one coupling
to quarks and the other coupling to leptons only.}  At least two of our
extra dimensions will always parameterize a $S^2$ with a magnetic
monopole like configuration on it. It has been shown in
\cite{sss1} that in such backgrounds the components of the gauge
field tangent to $S^2$ contain tachyonic modes. These are the
modes which we would like to interpret as the $D=4$ Higgs fields.
We shall give a simple general rule to identify the potential
tachyonic modes for any gauge group. We shall then work out three
examples. The first one will have a $D=6$ gauge group of $SU(3)$
with a triplet of fermions in its fundamental representation.  We
shall show that this leads to a $D=4$ effective theory with the
gauge group $SU(2)_L\times SU(2)_R \times U(1)$ with the fermions
in $(2, 1) + (1,2)$ representation of $SU(2)_L\times SU(2)_R$, so
that it can be interpreted as a left right symmetric model of
electroweak interaction of leptons. We shall work out the $D=4$
effective action for this model and the Yukawa couplings as well
as the
Higgs potential.\\

The second example will be in  $D=10$ for which we shall take the
gauge group to be $U(6)$ with a multiplet of $D=10$ chiral
fermions in the $\underline{6}$ of $U(6)$. The $U(6)$ Yang--Mills
equations can be solved by a self dual $U(1)$ instanton\footnote{
This instanton background on ${\mathbb{C}}P^2$ is necessary
in order for the spinors to be globally well defined on
${\mathbb{C}}P^2$. It defines the so called spin$^c$
structure on ${\mathbb{C}}P^2$.}
on
${\mathbb{C}}P^2$ and a magnetic monopole on
${\mathbb{C}}P^1=S^2$. The Kaluza--Klein gauge group will thus be
$SU(2)\times SU(3)$ We shall analyze the spectrum of the Dirac
operator on this background and identify the correct standard
model candidates for a single family of leptons and quarks. Our
general rule for the tachyonic contribution of the $S^2$
dependence will help us to identify the various candidates for
the Higgs fields among the components of the gauge field
fluctuations. However, these fluctuations now will also depend on
the ${\mathbb{C}}P^2$ coordinates. It is necessary to ensure that
upon harmonic expansion on ${\mathbb{C}}P^2$ the resulting modes
are singlets of $SU(3)$. We shall show that the Higgs tachyons
which give masses to the leptons and quarks are in fact singlets
of $SU(3)$.  Apart from this there are other fields whose masses
receive tachyonic contribution from $S^2$. These are potentially
dangerous and the leading term in their harmonic expansion on
${\mathbb{C}}P^2$ is a triplet of $SU(3)$. We need to ensure that
the ${\mathbb{C}}P^2$ contribution to their masses overwhelms the
tachyonic contribution of $S^2$. We show that the condition for
this to happen is that the ratio of the radii of $S^2$ and
${\mathbb{C}}P^2$ is bigger than $3/2$.  There are two ways to
ensure this, namely, either we give up the background Einstein
equations  in which case the two radii can be varied
independently, or we introduce an extra field in $D=10$ which
allows us to disentangle the two radii. There is not much to say
about the first solution.  As for the second one  we shall show
that the introduction of a $D=10$, $U(1)$  gauge field which
couples only to gravity solves our problem. Thus, as far as we
neglect the gravitational effects this gauge field will be
unobservable.\\

Finally we shall consider an  $U(N)$ model in $10$ dimensions in
which by an appropriate choice of the magnetic charges the tree
level masses of the would be tachyons will be zero. We shall
argue that the one loop induced mass can be made tachyonic. This
will provide us with the mild hierarchy which should exist
between the scale of electroweak symmetry breaking
and the compactification radius.\\

In this paper our intention is not to recover the standard model
of particle physics from a $10$-dimensional theory. In fact we
believe that within our present understanding this is not
possible.
\footnote{
For a review of attempts at model building using Kaluza--Klein
ideas, see \cite{zt}.}
Our aim is rather to see how close one can get to the
standard model if one insists in obtaining all of the three basic
ingredients, namely the chiral spectrum of fermions and Higgs
scalars which couple to the left handed doublets and the right
handed singlets and trigger the spontaneous breaking of gauge
symmetries. Although we obtain a spectrum of massless chiral
fermions and Higgs representation very similar to the one of the
standard electroweak-color theory our construction can not be
considered as realistic. Unless the tachyonic mass is induced by
a loop effect as we discuss in section
\ref{massless}, the masses of leptons
and the quarks which are produced by symmetry breaking are of the
same order as the masses of low lying  Kaluza--Klein modes which
we are discarding. Also if we take the radii of the internal
spaces to be of the order of TeV the masses of the leptons and
quarks will be of the same order. Our particles, if they are to
be identified with the standard model particles, should possibly
correspond to the third generation. Secondly, apart from the
$SU(2)\times SU(3)$ factor of the gauge group which we are
obtaining from the isometries of $S^2$ and ${\mathbb{C}}P^2$, the
original $U(6)$ has a $(SU(2)\times U(1))\times
(SU(2)\times U(1))\times
U(1)\times U(1)'$ factor, where the $U(1)'$ is generated by the
$6\times 6$ unit matrix,  which is also unbroken. Our putative
quarks and leptons transform non trivially under this group. If
we could make the couplings of this group week
of course there would be no problem. \\

In principle within the framework of the scheme introduced in
this paper, one could search for a more realistic group. However,
until our idea on the one loop generated tachyon mass  has been
put on a firmer ground,
the problem of scales would still persist.\\

The plan of this paper is as follows. In section \ref{2} we give
the background solution and discuss briefly its geometry. In
section \ref{section3} we discuss the fermion zero modes on $S^2$
and ${\mathbb{C}}P^2$. In section \ref{section4} we give our rule
for the tachyonic contribution of the $S^2$ factor to the masses
of scalars. In section \ref{examples} we discuss the example of
$G=SU(3)$ in $D=6$ and $G=U(6)$ in $D=10$. In section
\ref{section6} we analyze the ${\mathbb{C}}P^2$ contribution to
the masses of the Higgs doublet and give the condition for the
absence of $SU(3)$ non singlet tachyons. In section
\ref{section7} we analyze the $D=4$ scalars originating from the
gauge field fluctuations tangent to ${\mathbb{C}}P^2$ and show
that they are non-tachyonic.  In section \ref{massless} we study a
$D=10$ model with the gauge group $U(N)$ and show that, with a
particular choice of magnetic charges, the would be tachyons
become massless. In section \ref{summary}
we summarize the paper.\\

It should be noted that throughout the paper we discard the
scalar fields which originate from the gravity sector. These
scalars would not mix with the ones which we retain.\\

It should also be remarked that our examples  can suffer from
chiral anomalies. In our $10$ dimensional example they can be
removed by standard methods.  We comment on this in sections
\ref{anomaly} and \ref{summary}.


\section{The Background Solution\label{2}}


Although the background we are going to use will solve the field equations
of any generally covariant and gauge invariant action containing the metric
and the Yang--Mills fields only, for the sake of simplicity we
start from  Einstein--Yang--Mills system in
$D$-dimensions. The action is given by
$$
S= \int d^D x \sqrt{-G} \left(\frac{1}{\kappa^2}
{\cal R} -\frac{1}{2g^2}\mbox{Tr}F^2 +\lambda +
{\bar \psi } i \slash{\nabla} \psi \right )
$$
\nd
where $\psi$ is in some representation of the gauge
group $G$. This action can be the low energy string
field theory action with the $\lambda$-term induced
by some mechanism. The presence of $\lambda$ in our
discussion is required if we insist on having product
spaces like $M_1\times M_2\times...$ as a solution of
the classical bosonic field equations, where one of
the factors in the product is flat, e.g. the flat
$4$-dimensional Minkowski space. Our argument about
chirality is not sensitive to the flatness of any of
the factors in the product. The presence of tachyons,
however, depends on the definition of a mass
operator. This is different for example in AdS$^d$
and (Minkowski)$^d$. \\

The bosonic field equations are
$$
\frac{1}{\kappa^2}{\cal R}_{MN}=\frac{1}{g^2}
\mbox{Tr}F_{MR}{F_N}^R-\frac{1}{D-2} G_{MN}
\left(\frac{1}{2g^2}\mbox{Tr}F^2 + \lambda \right)
$$
$$
\nabla_MF^{MN}= 0
$$

In this paper we shall consider solutions of the form
$M_4\times K$, where $M_4$ is the flat $4$-dimensional
Minkowski space and $K$ is a compact manifold. In this
paper $K$ will be mostly taken to be either $S^2$ or
$S^2\times {\mathbb{C}}P^2$. Furthermore we shall
assume that the gauge field configuration $A$ will be
non-vanishing only on $K$. One can of course think of
many other choices for $K$.\\

The flatness of the Minkowski space implies
\begin{eqnarray}
&&\frac{1}{2g^2}\mbox{Tr}F^2 + \lambda=0\nonumber\\
&&{\cal R}_{{\hat m}{\hat n}}=\frac{\kappa^2}{g^2}
\mbox{Tr}F_{{\hat m}{\hat r}}{F_{\hat n}}^{\hat r}
\label{ink}
\end{eqnarray}
\nd
where $\hat m$, $\hat n$ are indices in $K$. Our
problem is now to find solutions of Yang--Mills
equations in $K$ which also solve the Einstein
equation (\ref{ink}).\\

For $K={\mathbb{C}}P^1\times {\mathbb{C}}P^2$ the
metric is given by
\begin{equation}
ds^2=a_1^2\left(d\theta^2+\mbox{sin}^2\theta d
\varphi^2 \right) +\frac{4a_2^2}{1+\zeta^{\dag}\zeta}
d{\bar \zeta}^a \left(\delta^{ab}-
\frac{\zeta^a{\bar\zeta}^b}{1+\zeta^{\dag} \zeta}
\right)d\zeta^b
\label{fub}
\end{equation}
\nd
where $a_1$ and $a_2$ are the radii of
${\mathbb{C}}P^1$ and ${\mathbb{C}}P^2$ respectively,
and $\zeta=(\zeta^1,\zeta^2)$ is a pair of local complex
coordinates in ${\mathbb{C}}P^2 $. The ${\mathbb{C}}P^2$
metric is the standard {\it Fubini-Study} metric. There
are two facts about ${\mathbb{C}}P^2 $ which are of
importance for our present discussion. The first is the
isometry group $SU(3)$ of ${\mathbb{C}}P^2 $. Together
with the invariance group $SU(2)$ of the metric of $S^2$,
$SU(3)$ will form part of the gauge group in $M_4$.
$SU(3)$ will be identified with the strong interaction
color gauge group. The low energy $4$-dimensional gauge
group will be ${\tilde G}\times SU(2)\times SU(3)$, where
$\tilde G$ is the subgroup of the $D$-dimensional gauge
group $G$ which leaves the background solution invariant.
Note that even with $G=U(1)$ we can obtain a
$4$-dimensional gauge theory with a gauge group $U(1)
\times SU(2)\times SU(3)$. Although such a solution can
produce chiral fermions in a non-trivial representation
of $U(1)\times SU(2)\times SU(3)$, it is not possible,
however,  to obtain the correct Standard Model spectrum
of leptons, quarks, and the Higgs fields. For this we
need a bigger $G$. We shall discuss this point in a
greater detail in a later section. \\

The second important fact about ${\mathbb{C}}P^2 $ is
that in the absence of a background $U(1)$ gauge field
it is not possible to have globally well defined spinor
field on it. This is principally due to the fact that
the complex coordinates $\zeta$ do not cover ${\mathbb{C}}
P^2 $ globally. We need at least three patches $(U,\zeta)$,
$(U',\zeta')$, and $(U{''},\zeta{''})$, where in $U\bigcap
U'$ we have the transition rule $\zeta_1'=
\frac{1}{\zeta_1}$ and $\zeta_2'=\frac{\zeta_2}{\zeta_1}$.
It needs some work to show that the
two chiral spinors of the
tangent space $O(4)$ of ${\mathbb{C}}P^2 $ can not be
patched consistently on the overlap. We shall give some more
details of this later on.\\

To write the solution of the Yang--Mills equations on
$K={\mathbb{C}}P^1\times {\mathbb{C}}P^2$ we first work out
the spin connection on $K$. It is given by
\begin{equation}
\Omega = -(\mbox{cos}\theta -1)d\varphi\frac{\tau^3}{2}+
\left(\begin{array}{cc}\frac{1}{2}\omega^i\sigma^i& 0\\
0& -\frac{3}{2}\omega \sigma_3\end{array}\right)
\label{X}
\end{equation}
\nd
where the first factor refers to ${\mathbb{C}}P^1$and the
second, which is a $4\times 4$ matrix, refers to
${\mathbb{C}}P^2$. Here $\tau^3$ as well as $\sigma^i$ and
$\sigma_3$ are Pauli matrices. Also the expressions are
valid on the upper hemisphere on ${\mathbb{C}}P^1$ and the
local patch $(U,\zeta)$ on ${\mathbb{C}}P^2$. The expressions
for $\omega^i$ and $\omega$ can be read from the Fubini-Study
metric (\ref{fub}) on ${\mathbb{C}}P^2$. We shall not need
the explicit expression for $\omega^i$. The one for $\omega$
is given by
\begin{equation}
\omega(\zeta, {\bar \zeta})= \frac{1}{2\left(1+\zeta^{\dag}
\zeta\right)}\left(\zeta^\dag d\zeta - d\zeta^\dag \zeta
\right)
\label{a}
\end{equation}
\nd
Note that $d\omega$ is the self dual K\"{a}hler form on
${\mathbb{C}} P^2$. It is thus an instanton type solution of
the Yang-Mills equation in ${\mathbb{C}} P^2$.\\

It is important note from (\ref{X}) that
the
${\mathbb{C}} P^2$ spin-connection takes its values in the
subgroup $SU(2)
\times U(1)$ of the tangent space $SO(4)$. Furthermore, under
$ SO(4)\rightarrow SU(2)\times U(1) $ the two chiral spinors
of $O(4)$ decompose according to
\begin{eqnarray}
2_+&=& 2_0 \\
2_-&=&1_{-\frac{3}{2}}+1_{\frac{3}{2}}
\end{eqnarray}
\nd
where the subscripts indicate the $U(1)$-charges. Using this
fact one can understand why spinors are not globally well
defined on ${\mathbb{C}}P^2$. The point is that in the overlap
of two patches $(U,\zeta)$ and $(U',\zeta')$ we have
\begin{equation}
\omega(\zeta')= \omega(\zeta) -id\varphi
\end{equation}
\nd
where $\varphi$ is defined by $\zeta_1=|\zeta_1|
\mbox{e}^{i\varphi}$. For $2_-$ to be globally well defined
$1_{\pm 3/2}$ should patch according to the rule $\psi'
(\zeta')=\mbox{e}^{\pm \frac{3}{2}i\varphi} \psi(\zeta)$. We
thus obtain transition functions which are anti-periodic under
$\varphi\rightarrow \varphi + 2\pi$. Coupling a background
gauge field proportional to $\omega$ can change this. With a
little more work one can show that a similar obstruction also
prevents $2_+=2_0$ from being well defined.\\

Now we are in a position to write our solution of the
Yang--Mills equation on  ${\mathbb{C}}P^1\times {\mathbb{C}}
P^2$. It is easy to show that the ansatz
\begin{equation}
A=\frac{n}{2}(\mbox{cos}\theta
-1)d\varphi + qi\omega
\label{back}
\end{equation}
\nd
where $n=\mbox{diag}(n_1, n_2,...)$ and $q=\mbox{diag}(q_1,
q_2,...) $ are matrices in the Cartan-subalgebra of $G$. The
consistent patching of spinors requires that $n_1, n_2,...$
be integers and $q_1,q_2,...$ be one half of an odd integer.
Note that the substitution of the above ansatz in the Einstein
equations will require that the radii $a_1$ and $a_2$ of
${\mathbb{C}}P^1$ and ${\mathbb{C}}P^2$ are quantized.\\

As mentioned in the beginning
of this section our ansatz for the background
configuration solves the field equations derived from any
generally covariant and gauge invariant Lagrangian in $D=10$, which contains
the metric and the Yang-Mills potentials only. Such an effective Lagrangian
will contain infinite number of parameters and therefore the relationship
between the radii and other parameters will be more involved.


\section{Chiral Fermions\label{section3}}


It is a well known fact that in order to obtain chiral
fermions in $D=4$ we need topologically non-trivial
background
gauge fields on
${\mathbb{C}}P^1\times{\mathbb{C}}P^2$. Our solution for the
Yang--Mills equations consist of magnetic monopole on $S^2$
and the potential for the K\"{a}hler form on ${\mathbb{C}}P^2$.
The  K\"{a}hler form defines a topologically non trivial line bundle on
${\mathbb{C}}P^2$ . \\

Consider the $D=10$ fermion Lagrangian
\begin{equation}
{\cal L}={\bar \psi}i \slash{\nabla}\psi
\end{equation}
\nd
where
\begin{equation}
\nabla_{\hat M}\psi = (\partial_{\hat M}+\omega_{\hat M}
-i A_{\hat M}
)\psi\;\;, \;\; {\hat M}=0,1,...,9
\end{equation}
\nd
$\omega_{\hat M}$ and $A_{\hat M}$ are, respectively, the
$SO(1,9)$ and the Lie algebra valued spin and gauge connections.
We analyze the fermion problem in two steps. In the first step
we write the manifold as $M_6\times  {\mathbb{C}}P^2 $.
Correspondingly we write the $D=10$ Dirac matrices as
\begin{eqnarray}
{\hat \Gamma}_a& =&\Gamma \times \gamma_a\;\;\;\;\;\;\;a=6,7,
8,9\nonumber\\
{\hat \Gamma}_A &=&\Gamma_A \times 1\;\;\;\;\;\;\;A=0,1,...,
5\nonumber
\end{eqnarray}
\nd
where $\gamma_a$ and $\Gamma_A$ are respectively $4\times 4$
and $8\times 8 $ Dirac matrices satisfying
\begin{eqnarray}
\{\gamma_a,\gamma_b \}&=&2\delta_{ab} \nonumber\\
\{\Gamma_A,\Gamma_B \}  &=&2\eta_{AB}\nonumber
\end{eqnarray}
\nd
and $\Gamma=\Gamma_0\Gamma_1...\Gamma_5$.\\

Substituting these $\Gamma$'s into ${\cal L}$ and recalling
that the geometry has factorized form we obtain
\begin{equation}
{\cal L}={\bar \psi}\Gamma i \slash{\nabla}_{ {\mathbb{C}}P^2
}\psi + {\bar \psi}i \slash{\nabla}_{M_6 }\psi
\end{equation}
\nd
The chiral fermions on $M_6$ will originate from those modes
for which
\begin{equation}
\slash{\nabla}_{ {\mathbb{C}}P^2
}\psi=0
\label{psi}
\end{equation}
\nd
Those $\psi$'s which are not annihilated by
$ \slash{\nabla}_{{\mathbb{C}}P^2}$ will give rise to massive
fermionic modes on $M_6$. The standard way to analyze
(\ref{psi}) is to operate one more time with
$ \slash{\nabla}_{{\mathbb{C}}P^2}$ on it. Using the background
connections (\ref{X}) and $(\ref{back})$ we obtain
\begin{equation}
(\nabla^2-\frac{3}{2})\;\psi_+=0
\label{psi+}
\end{equation}
\begin{equation}
\{ \nabla^2+(q\;\sigma_3-\frac{3}{2}) \}\;\psi_-=0
\label{-}
\end{equation}
\nd
where
\begin{equation}
\nabla \psi_+=(d+i\omega^r\frac{\sigma^r}{2}+\omega\;q)\;\psi_+
\end{equation}
\begin{equation}
\nabla \psi_-=\{d+\omega(q-\frac{3}{2}\sigma_3)\}\;\psi_-
\end{equation}
\nd
and
$$
\psi_\pm =\frac{1\pm {\hat \gamma}_5}{2}\;\psi\;\;\;\;,\;\;\;\;
{\hat \gamma}_5 =\gamma_6\gamma_7\gamma_8\gamma_9$$
\nd
The K\"{a}hler instanton $\omega$ is given by equation
(\ref{a}). Since
$\nabla^2\leq 0$ (\ref{psi+}) will have no non-zero solutions.
Thus fermions of $\psi_+$ type will all be non-chiral and massive.
Equation (\ref{-}), on the other hand, can have solutions. Their
existence depends on the eigenvalues of $q$. Clearly for $q=3/2$
we have only one solution with $\sigma_3=+1$. For $q=+5/2$ we
obtain $3$ solutions with $\sigma_3=+1$. They form a triplet of
the isometry group $SU(3)$ of ${\mathbb{C}}P^2$. For $q=-5/2$ and
$\sigma_3=-1$ we obtain a $3^*$ of $SU(3)$. These are the only type
of solutions we need to consider. \\

Next we study the $M_6$ Dirac Lagrangian
\begin{equation}
{\cal L}={\bar \psi}i \slash{\nabla}_{M_6 }\psi
\label{6}
\end{equation}
\nd
where $\psi$ is assumed to be a solution of (\ref{-}). We shall
assume that the $D=10$ spinor is chiral and has positive
chirality. Then the spinor of $\psi_-$ type will have
negative
$D=6$ chirality. We choose the $D=6$ $\Gamma$ matrices to be
\begin{eqnarray}
\Gamma_\alpha& =&\Gamma_\alpha \times \tau_1\;\;\;
\;\;\;\;\alpha=0,1,2,3\nonumber\\
\Gamma_4 &=&\Gamma_5 \times \tau_1\;\;\;\;\;\;\;
\gamma_5=i\gamma_0\gamma_1\gamma_2\gamma_3\nonumber\\
\Gamma_5&=&1\times \tau_2
\end{eqnarray}
\nd
and $\tau_1= \left(\begin{array}{cc}
0&1 \\
1& 0\end{array}\right)$,  $\tau_2= \left(\begin{array}{cc}
0&-i \\
i& 0\end{array}\right)$.\\

Inserting the $\Gamma_A$'s in (\ref{6}) we obtain
\begin{equation}
{\cal L}={\bar \psi}i \slash{\nabla}_{M_6 }\psi
+\frac{i}{\sqrt{2}} \left\{{\bar \psi}(\gamma_5+1)D_-
\psi+ {\bar \psi}(\gamma_5-1)D_+\psi
\right \}
\end{equation}
\nd
where
\begin{equation}
D_\pm \psi=e_{\pm}^m\left(\partial_m +\frac{i}{2}\omega_m
(n-\gamma_5)
\right)\psi
\end{equation}
\nd
$e_{\pm}^m$ are the $U(1)$ components of an orthonormal frame
on $S^2$ and $\omega_m$ is the corresponding spin connection
($\omega_\theta =0$, $\omega_\varphi=-\mbox{cos}\theta+1$
in the upper hemisphere and
$\omega_\varphi=-\mbox{cos}\theta-1$ in the lower hemisphere
).\ Decomposing
$\psi=\psi_L+\psi_R$, where
$\displaystyle\psi_L=
\frac{1-\gamma_5}{2} \psi$,\ we obtain the analogue of
(\ref{psi+}, \ref{-}) for the Dirac operator on ${\mathbb{C}}
P^1$
$$
\left\{\nabla^2-\frac{1}{2}(1-n)  \right\}\psi_R=0
$$
$$
\left\{\nabla^2-\frac{1}{2}(1+n)  \right\}\psi_L=0
$$
\nd
$n=1$ produces one $\psi_R$ while $n=-2$ gives rise to two
$\psi_L$ which form a doublet of the Kaluza--Klein $SU(2)$.


\section{General Rule for Higgs Type Tachyons\label{section4}}


To obtain the spectrum of the effective theory in
$4$-dimensions we need to expand the functions about our
background solution in harmonics on ${\mathbb{C}}P^1\times
{\mathbb{C}}P^2$. These include fluctuations of the
gravitational, Yang--Mills, as well as fermionic fields. The
techniques of doing such analysis have been developed long
ago. In this paper we shall ignore the gravitational
fluctuations and consider only the Yang--Mills and fermionic
fields. The full set of linearized gravity Yang--Mills
equations can be found in \cite{sss1}. In the same paper it
was shown that there are tachyonic modes in the components of
the gauge field fluctuations tangent to $S^2$. Here, we would
like to show that the rule to identify the tachyonic modes
given in \cite{sss1} for $G\equiv SU(3)$ is in fact quite
general and applies to any gauge group $G$. It should be
emphasized that neglecting the gravitational fluctuations is
justified as they will not mix with the gauge field
fluctuations of interest for us. \\

In general, we should write $A={\bar A}+V$ where $\bar A$ is
the background solution and $V$ depends on the coordinates of
$M_4$, $S^2$ and ${\mathbb{C}}P^2$. Our first interest is in
the fields which are tangent to $S^2$. It is these fields,
which if develop a tachyonic vacuum expectation value, can
break $SU(2)$, provided such modes are singlets of $SU(3)$
isometry of ${\mathbb{C}}P^2$.\\

We suppress the ${\mathbb{C}}P^2$ dependence of these fields
and denote by $V_{\underline 1}$ and $V_{\underline 2}$ their
components with respect to an orthonormal frame on $S^2$. It
is convenient to use the ``helicity'' basis on $S^2$ defined
by
$$
V_{\pm}=\frac{1}{\sqrt{2}}(V_{\underline 1} \mp i\;
V_{\underline 2})
$$
\nd
$V_{\pm}$ are matrices in the Lie algebra of $G$. What governs
their mode expansion on $S^2$ is their isohelicities. This is
basically the effective charge of $V_{\pm}$ under the
combination of $U(1)$ transformations which leave our background
configuration invariant. These charges can be evaluated in the
same way which was done in \cite{sss1}. For the sake of
simplicity, let us assumes $G=U(N)$ and assume that charge
matrices $n$ and $q$ introduced in (\ref{back}) are diagonal
$N\times N$ matrices. Then $V_\pm$ are $N\times N$ matrices with
elements ${V_{\pm i}}^j$, $i,j=1,...N$. Their isohelicities,
$\lambda({V_{\pm i}}^j)$, are given by
$$
\lambda({V_{\pm i}}^j)=\pm 1 + \frac{1}{2}(n_i-n_j)
$$
\nd
Note that there is a hermiticity relation
$$
{V_{+i}}^j = \left( {V_{-j}}^i \right)^*
$$
\nd
The harmonic expansion of ${V_{+i}}^j$ on $S^2$ will produce an
infinite number of Kaluza--Klein modes. These expansions are
defined by \begin{equation}
V_\pm (x,\theta,\varphi)= \sum_{l\geq |\lambda_\pm|}
\sqrt{\frac{2l+1}{4\pi}} \sum_{m\leq|l|} V_\pm^{lm} (x)\;
D_{\lambda_\pm,m}^l(\theta,\varphi)
\end{equation}
\nd
$ D_{\lambda_\pm,m}^l(\theta,\varphi)$ are $2l+1$-dimensional
unitary matrices.\\

The tachyonic modes are generally contained in the leading terms
with $l=|\lambda_\pm|$. The effective $4$-dimensional mass$^2$
of $V_\pm^{lm}(x)$ obtains contributions from the appropriate
Laplacian acting on $S^2$ and ${\mathbb{C}}P^2$. $V_\pm$ are
charged scalar fields on ${\mathbb{C}}P^2$. We shall analyze
their dependence on the ${\mathbb{C}}P^2$
coordinates in the next section.
Here we shall consider the $S^2$ contribution to their masses.
The condition for this contribution to be tachyonic is expressed
in the following simple rule
$$
M^2({V_{+i}}^j) < 0\;\;\;\;\mbox{if}\;\;\;\;\lambda( {V_{+i}}^j)
\leq 0$$
\nd
Likewise
$$
M^2({V_{-i}}^j) < 0\;\;\;\;\mbox{if}\;\;\;\;\lambda( {V_{-i}}^j)
\geq 0$$
\nd
To prove these claims let us make more detailed analysis. \\

Since we are assuming $V_\pm$ are independent of the
${\mathbb{C}}P^2$ coordinates, their mass term comes from the
expansion of $\mbox{Tr} F_{mn}F^{mn}$, where $m,n$ indicate
indices tangent to $S^2$. The cubic and the quadratic parts in
$\mbox{Tr} F_{mn}F^{mn}$ will produce the interaction terms in
the Higgs potential. We have
\begin{eqnarray}
\mbox{Tr}
F_{mn}F^{mn}&=&  \mbox{Tr}{\bar F}_{mn}{\bar F}^{mn} +
\mbox{Tr}(D_+V_- - D_-V_+)^2\nonumber\\
&-&4\;i\;  \mbox{Tr}{\bar F}_{+-} [V_-,V_+]
-2\;i \;\mbox{Tr}(D_+V_- - D_-V_+) [V_-,V_+]\nonumber\\
&+&\mbox{Tr}[V_-,V_+]^2
\nonumber
\end{eqnarray}
\nd
where the covariant derivatives are defined by
$$
D_mV_n=\nabla_m V_n - i[{\bar A}_m,V_n]
$$
\nd
$\nabla_m$ denotes the ordinary Riemannian covariant
derivative on $S^2$. Now, since for $\lambda_+\leq 0$ \
($ \lambda_+\geq 0 $)
$D_- D_{\lambda_+,m}^{l=|\lambda_+|}=0  =  D_+
D_{\lambda_-,m}^{l=|\lambda_-|}   $
we see that such modes will be annihilated by $D_\pm$ and thus
the $S^2$ contribution to their $D=4$ action is given by
$$
S=-\frac{1}{2g^2}\int_{0}^{2\pi}d\varphi\int_{0}^{\pi}
\mbox{sin}\theta \;\mbox{Tr}\left\{ 4D_\mu V_+D^\mu V_- -4i
{\bar F}_{+-} [V_-,V_+]
+[V_-,V_+]^2 \right\}
$$
\nd
The mass terms hence come from $-4iTr{\bar F}_{+-}[V_-,V_+]$
term only. \\

To proceed it is convenient to choose the Cartan-Weyl basis for
the Lie algebra of $G$. Let $Q_j$ denote the basis of the Cartan
subalgebra, $E_\alpha$ and $E_{-\alpha}=E_\alpha ^\dag $ the
generators outside the Cartan subalgebra. The only part of the
algebra needed for the evaluation of the mass terms is
$$
[Q_j,E_\alpha]= \alpha_j E_\alpha
$$
\nd
In this basis we can write
$$
V_\pm= V_\pm^\alpha E_\alpha+ (V_\mp^\alpha)^*E_{-\alpha}
+V_\pm^jQ_j
$$
\nd
It is easy to see that
\begin{equation}
\lambda(V_\pm^\alpha)=\pm 1+ p.\alpha
\end{equation}
\nd
where $p.\alpha=p^j\alpha_j $ and $p^j$ are defined by
$$
\frac{1}{2}n=p^jQ_j
$$

To
simplify the discussion consider the case when only one
$\lambda(V_+^\alpha)\leq 0$. Set the remaining modes to zero.
Of course this is not a loss of generality. In this case
$V_+=V_+^\alpha E_\alpha$ and
$V_-=(V_+^\alpha)^*E_{-\alpha}$. The mass term then becomes
\begin{eqnarray}
\mbox{Tr}\left(-4i {\bar F}_{+-}[V_-,V_+]\right)
&=&-4i
\mbox{Tr} V_+ [ {\bar F}_{+-} ,V_-]
\nonumber\\
&=& \frac{4}{a_1^2}\;p.\alpha\;|V_+^\alpha|^2
\mbox{Tr}E_\alpha E_{-\alpha}
\nonumber
\end{eqnarray}
\nd
where we inserted $\displaystyle {\bar F}_{+-}=-
\frac{i}{a^2}p^jQ_j$. The kinetic part of the action for
$V_+^\alpha$ thus becomes
$$
S_2= -\frac{2Tr( E_\alpha E_{-\alpha})}{g^2}
\int_{0}^{2\pi}d\varphi\int_{0}^{\pi} d\theta \;
\mbox{sin}\theta \left\{
\partial_\mu V_+^\alpha\partial^\mu {V_+^\alpha)}^*+
\frac{p.\alpha}{a_1^2}{|V_+^\alpha|^2}
\right\}
$$

Substituting $p.\alpha=\lambda(V_+^\alpha)-1$ we obtain the
mass of $V_+^\alpha$ in terms of its isohelicity as (recall
that our signature is $(-,+,+,...)$)
\begin{equation}
m^2=\frac{\lambda-1}{a_1^2}
\label{mass}
\end{equation}
\nd
which is negative for $\lambda\leq 0$.
Similar reasoning can be applied if for some $V_-^\alpha$
the corresponding isohelicity $\lambda(V_-^\alpha) $ is
non-negative.\\

This rule gives us an easy way of identifying possible
tachyonic modes which can act as Higgs scalars in the $D=4$
effective theory.


\section{Examples\label{examples}}


In this section we shall ignore the
${\mathbb{C}}P^2$ part and give some examples of a $D=6$
gravity Yang--Mills theories which produce standard model
type Higgs sectors upon compactification to $D=4$. Leptons
and quarks will be included in the next sections.
We basically need to choose the gauge group  $G$ and assign
magnetic charges $n$. \\

The notation is always
\begin{equation}
{\bar A}=\frac{n}{2}(\mbox{cos}\theta\mp 1)d\varphi
\end{equation}
\nd
where $
n=\mbox{diag}(n_1,n_2,...)
$ is in the Lie algebra of $G$, $-(+)$ give the expression
for $\bar A$ in the upper (lower) hemispheres.\\


\subsection{Tachyons}


\subsubsection{\mbox{\boldmath $G=SU(3)$}}

\begin{equation}
n=\mbox{diag}(n_1,n_2,-n_1-n_2),\;\;\;\;\;\;n_1,n_2\in {\mathbb{Z}}
\end{equation}
\nd
The isohelicities can be assembled in a $3\times 3$ matrix
\begin{equation}
\lambda(V_\pm)=
\left(\begin{array}{ccc}
\pm 1 & \pm 1+\frac{1}{2}( n_1-n_2)& \frac{1}{2}( 2n_1+n_2) \\
\pm 1- \frac{1}{2}(n_1-n_2) &\pm 1 &\pm 1 +\frac{1}{2}(n_1+2n_2) \\
\pm 1-\frac{1}{2}(2n_1+n_2) & \pm 1-\frac{1}{2} (n_1+2n_2)& \pm 1
\end{array}
\right)
\end{equation}

Using the results of
section
\ref{section3} we see that in order to obtain
left handed doublets and right handed singlets we had to
take $(n_1,n_2)=(1,1)$. With these values of $n_1$ and $n_2$,
${V_{-1}}^3$ and ${V_{-2}}^3$ will contain tachyonic modes
in the leading term of their expansion on $S^2$.\\

In this example the  $SU(2)\times U(1)$ subgroup of $SU(3)$
is unbroken and the tachyonic Higgs ${V_{-1}}^3$ and
${V_{-2}}^3$ form a doublet of $SU(2)$ with $U(1)$ charge
of $3/2$. We denote this doublet by $\phi$. Its isohelicity
is $+1/2$. Therefore it will also be a doublet of the
Kaluza--Klein isometry of $S^2$. One can integrate the
$(\theta,\varphi)$ dependence of $\phi$ on $S^2$ and work
out its $D=4$ effective action. The result is
$$
{\cal L}= -\frac{1}{2g^2}
\int_{0}^{2\pi}d\varphi\int_{0}^{\pi} d\theta \;
\mbox{sin}\theta \;\mbox{Tr} F_{MN}F^{MN}
\;\;\;\;\;\;\;\;\;\;\;
\;\;\;\;\;\;\;\;\;\;\;
\;\;\;\;\;\;\;\;\;\;\;$$
$$
\;\;\;\;\;
= -\frac{1}{4g_1^2} {F^8_{\mu\nu}}^2 -
\frac{1}{4g_2^2} {F^r_{\mu\nu}}^2
-\frac{1}{4e^2} {W^r_{\mu\nu}}^2
$$
\begin{equation}
\;\;\;\;\;\;\;\;\;\;\;\;
\;\;\;\;\;\;\;\;\;\;\;
\;\;\;\;\;\;\;\;\;\;\;
\;\;\;\;\;\;\;\;
-\mbox{Tr}\left\{
\nabla_\mu \phi^\dag \nabla^\mu \phi -\frac{3}{2a_1^2}
\phi^\dag\phi+2g_1^2 (\phi^\dag\phi)^2\right\}
\end{equation}
\nd
where we have regarded $\phi$ as a $2\times 2$ complex
matrix, and
$$
\nabla_\mu \phi=\partial_\mu\phi-\frac{3}{2}iV_\mu^8\phi
-iV_\mu^r \frac{\sigma^r}{2}\phi-i W_\mu^r \phi
\frac{\tau^r}{2}$$
\nd
where $V_\mu^8$, $V_\mu^r$, and $W_\mu^r$ are
respectively the $U(1)$, $SU(2)_L$, and the Kaluza--Klein
$SU(2)_R$ gauge fields. $g_1$, $g_2$, and $e$ are their
respective couplings. Some calculation show that
\begin{equation}
g_2=\frac{1}{2\sqrt{\pi}} \frac{g}{a_1}=\sqrt{3} g_1
\end{equation}
\nd
The Kaluza--Klein gauge coupling $e$ can also be expressed
in terms of the fundamental scales $g$ and $a_1$.\\

In the next section we shall work out the Yukawa couplings
for this model as well.

\subsubsection{ \mbox{\boldmath $G=U(6)$} }

With $n=\mbox{diag}(n_1,..., n_5,n_6)$
we can again work
out the table of isohelicities for $V_\pm$. We shall see in
section \ref{5.2} that in order to obtain one family of leptons
and quarks we need to take $n=\mbox{diag}(-2,1, 1,-2,1,1)$.
Note that since the group is $U(6)$ rather than $SU(6)$, $n$
is not traceless. $\lambda(V_\pm)$ is given by
\begin{equation}
\displaystyle
\lambda(V_+)=
\left(\begin{array}{cccccc}
+1
\;&-\frac{1}{2}
\;&-\frac{1}{2}
\;&+ 1
\;&-\frac{1}{2}
\;&-\frac{1}{2}
\\
+\frac{5}{2}
\;&+1
\;&+1
\;&+\frac{5}{2}
\;&+1
\;&+ 1  \\
+\frac{5}{2}
\;&+1
\;&+1
\;&+\frac{5}{2}
\;&+1
\;&+1
\\
+1
\;&-\frac{1}{2}
\;&-\frac{1}{2}
\;&+1
\;&-\frac{1}{2}
\;&-\frac{1}{2}
\\
+\frac{5}{2}
\;&+1
\;&+ 1
\;&+\frac{5}{2}
\;&+ 1
\;&+ 1
\\
+\frac{5}{2}
\;&+1
\;&+1
\;&+\frac{5}{2}
\;&+1
\;&+1
\end{array}
\right)
\label{iso}
\end{equation}
\nd
Since
$V_-=V_+^\dag$,
therefore $\lambda({V_{-i}}^j)=-\lambda({V_{+j}}^i)$.\\

The tachyonic modes are contained in ${V_{+1}}^i$,
and
${V_{+4}}^i$ where $i=2,3,5,6$.
They will all be doublets
of the Kaluza--Klein $SU(2)$. They also transform under some
representation of the unbroken part of $U(6)$, which is $SU(2)
\times SU(2)
\times U(1)^3\times U(1)'$, which is generated by
$\displaystyle
\mbox{diag}(0,\frac{\sigma^i}{2}, 0,0,0)$, \\
$i=1,2,3$;
$\displaystyle\mbox{diag}(0,0,0,0,\frac{\sigma^i}{2})$;
$\mbox{diag}(-2,1,1,0,0,0)$;
$\mbox{diag}(0,0,0,-2,1,1)$;\\
$\mbox{diag}(1,1,1,-1,-1,-1)$;
and the $6\times 6$ unit matrix
$1_6$ which generates $U(1)'$. The tachyonic Higgs will be
neutral under this $U(1)'$, therefore  their tree level
vacuum expectation value will not break it. Under
$U(6)\rightarrow SU(2)\times SU(2)\times U(1)^3$
we have
\begin{equation}
\underline{6}= (1,1)_{(-2,0,1)}
+(2,1)_{(1,0,1)}+ (1,1)_{(0,-2,-1)}+
(1,2)_{(0,1,-1)}
\end{equation}
\nd
The quantum numbers of the relevant Higgs tachyons will be
\begin{eqnarray}
{V_{+1}}^i &\sim& (2,1)_{(-3,0,0)}\;\;\;i=2,3 \label{v+1}\\
{V_{+4}}^t &\sim& (1,2)_{(0,-3,0)}\label{67} \;\;\;t=5,6
\end{eqnarray}

As we said earlier, the tachyonic modes in all these fields
will be in the doublet representation of the Kaluza--Klein
$SU(2)$. The vacuum expectation value of
the fields ${V_{+1}}^i\sim (2,1)_{(-3,0,0)}$ and
${V_{+4}}^t \sim
(1,2)_{(0,-3,0)}$ will give masses to the quarks and leptons
respectively. In section \ref{section6} we shall show that
the leading term in their expansion on ${\mathbb{C}}P^2$
is a singlet of $SU(3)$ and therefore their masses
receive no contribution from the dependence on the
${\mathbb{C}}P^2$ coordinates. Thus they remain tachyonic.
The
other tachyonic fields,
namely ${V_{+1}}^i$, $i=5,6$; ${V_{+4}}^t$, $t=2,3$,
would induce Yukawa couplings between
quarks and leptons.
We shall show that in fact the leading term in
their harmonic expansion on  ${\mathbb{C}}P^2$ is
a triplet of
$SU(3)$. Thus the vacuum expectation value
of these fields can
break the color $SU(3)$.
We will determine the conditions to avoid this.


\subsection{Fermions\label{anomaly}}


We consider the two examples of the previous section.

\subsubsection{
\mbox{\boldmath $G=SU(3)$} \label{anomaly1}}

Here we assume that $D=6$ and there is no ${\mathbb{C}}P^2$
factor. Let us take $\psi$ in $\underline{3}$ of $SU(3)$ and
$n=\mbox{diag} (1,1,-2)$. According to our rules this will
produce two right handed singlets of the Kaluza--Klein $SU(2)$
which we denote by $SU(2)_K$ and a left handed doublet. The
singlets will form a doublet of $SU(2)_G\subset SU(3)$ and the
doublet of $SU(2)_K$ will be a singlet of $SU(2)_G$. Thus under
$SU(2)_K\times SU(2)_G\times U(1)$ where $U(1)\subset SU(3)$ we
have $(1,2_R)_{1/2}+(2_L,1)_1$.
The $D=4$ Yukawa
and gauge
couplings can be easily worked out. The result is
\begin{eqnarray}
{\cal L}_F&=&
\int_{0}^{2\pi}d\varphi\int_{0}^{\pi}d\theta\;
\mbox{sin}\theta\; {\bar \psi}i \slash{\nabla}\psi
\;\;\;\;\;\;\;\;
\;\;\;\;\;\;\;\;
\;\;\;\;\;\;\;\;
\;\;\;\;\;\;\;\;
\;\;\;\;\;\;\;\;
\;\;\;\;\;\;\;\;
\nonumber\\
&=&{\bar \lambda}_L\; i \gamma^\mu\left(
\partial_\mu-ig_1V_\mu^8-ie W_\mu^i\frac{\tau^i}{2}
\right)\lambda_L
\nonumber\end{eqnarray}
$$
\;\;\;\;\;\;\;\;
+
{\bar \lambda}_R\; i \gamma^\mu\left(
\partial_\mu-i\frac{g_1}{2}V_\mu^8-ig_2 V_\mu^i
\frac{\sigma^i}{2}\right)\lambda_R
$$
$$\;\;\;\;\;\;\;\;
\;\;\;\;\;\;\;\;\;
\;\;\;
-2g_1
\left\{{\bar \lambda}_L\phi (i\sigma_2)\lambda_R
-{\bar \lambda}_R (i\sigma_2)\phi^\dag\lambda_L
\right\}
$$
\nd
where $\lambda_L=(2_L,1)_1$ and $\lambda_R=(1,2_R)_{1/2}$.\\

This expression together with the bosonic part given in
equation (\ref{6})
give the total effective $D=4$ action for the
$SU(3)$ example. Although this example leads to interesting
chiral and Higgs spectrum in $D=4$ can not be considered
satisfactory. It has both perturbative and global chiral
anomalies in $D=6$. The perturbative anomalies can be
eliminated with the standard Green Schwarz
mechanism\cite{gs}. To
apply this mechanism \cite{sss3} we need first to introduce an
antisymmetric rank two potential together with three right
handed
$D=6$ $SU(3)$ singlets to kill the pure
gravitational anomaly which is given by ${\cal R}^4$ term in the
anomaly $8-$ form. The remaining terms in the anomaly
$8$-form factorize appropriately in order to be canceled
by a judicious transformation of the antisymmetric
potential. This mechanism does not cancel the global
anomalies \cite{witten2}
whose presence is due to the fact that $\pi_6
(SU(3))=Z_6$ is non zero. To kill these ones we need to
introduce further $SU(3)$ multiplets or to change the gauge
group altogether and chose to a gauge  group
like
$E_6$ which has a trivial $\pi_6
(E_6)$.

\subsubsection{ \mbox{\boldmath $G=U(6)$}\label{5.2}}

Now assume $D=10$ and choose $\psi$ to be in $\underline{6}$ of
$U(6)$ and $q=\mbox{diag}(5/2,5/2,5/2,\\
3/2,3/2,3/2)$. As before
$n$ will  be taken to be $n=\mbox{diag}(-2,1,1,-2,1,1)$.\
According to the  results of the previous section with
respect to the isometry  group $SU(2)\times SU(3)$ we have
the following chiral fermions
$$
(2_L,\underline{3})+
(1_R,\underline{3})+
(1_R,\underline{3})+
(2_L,1)+
(1_R,1)+(1_R,1)
$$
\nd
Clearly the first three triplets are candidates for $\left(
\begin{array}{c}
u\\ d
\end{array}\right)_L$, $u_R$ and $d_R$. The last two pieces can
be identified with the leptons $\left(\begin{array}{c}\nu_e\\ e
\end{array}\right)_L$ and $e_R$.\footnote{
We have an extra right handed singlet in the lepton sector. This
can be removed by choosing the last entry in $q$ to be for
instance $-1/2$ or the last entry in $n$ to be $0$. In this
way the unbroken subgroup of $U(6)$ will be
$SU(2)\times U(1) \times U(1)'$.}
\\

These multiplets also transform in the following representation
of the unbroken $SU(2)\times SU(2)\times U(1)^3\subset G$
\begin{eqnarray}
(2_L,\underline{3})&\sim&(1,1)_{(-2,0,1)}\nonumber\\
(1_R,\underline{3})+(1_R,\underline{3})
&\sim& (2,1)_{(1,0,1)}\nonumber\end{eqnarray}
$$
(2_L,1)\sim  (1,1)_{(0,-2,-1)}\;\;\;\mbox{and}\;\;\;
(1_R,1)\sim  (1,2)_{(0,1,-1)}
$$
\nd
The Yukawa coupling between the quarks will be through the Higgs
field ${V_{+1}}^i$ given in (\ref{v+1}), while the electron will
get its mass through coupling to ${V_{+4}}^t$.  Thus our
construction leads to a multi Higgs theory in which the
quarks and leptons obtain their masses from their Yukawa
couplings to different Higgs scalars.  Note also that
there is no common $U(1)$ under which both Higgs
multiplets are charged. The hypercharge coupling in our
model is different from the standard electroweak theory.


\section{Higgs Like Tachyons on \mbox{\boldmath ${\mathbb{C}}P^2
\times {\mathbb{C}}P^1$}\label{section6}}


If the total space-time dimension is $D=6$ the masses of the Higgs
like tachyons are given by (\ref{mass}). In the case of a $D=10$
theory we need to take into account the contribution of
${\mathbb{C}}P^2$ part as well. The fields $V_\pm$ are like scalar
fields on ${\mathbb{C}}P^2$ which are
charged with respect to the
${\mathbb{C}}P^2$ part of the background gauge field (\ref{back}),
viz, $iq\omega$. The
${\mathbb{C}}P^2$
contribution to the masses of $V_\pm$
come
from the commutator term in the ${\mathbb{C}}P^2$ covariant
derivative of $V_\pm$, i.e.
\begin{eqnarray}
DV_\pm&=&dV_\pm-i[i \omega q, V_\pm] \nonumber\\
&=&dV_\pm+\omega[q,V_\pm]\nonumber
\end{eqnarray}
\nd
To be specific let us consider the example of the $U(6)$ model
for which $q=\mbox{diag}(5/2,5/2,5/2;3/2,3/2,3/2)$. Write
\begin{equation}
V=
\left(\begin{array}{c|c}
\;\;v\;\;&\;\;u\;\;\\
\hline
\;\;{\tilde u}\;\;& \;\;{\tilde v}\;\;\end{array}
\right)
\label{V}
\end{equation}
\nd
where $v$, $\tilde v$, $u$, and $\tilde u$ each is a
$3\times 3$ matrix.
Then
$$
[q,V]=
\left(\begin{array}{c|c}
\;\;0\;\;&\;\;u\;\;\\
\hline
\;-{\tilde u}\;\;&\;\; 0\;\;\end{array}
\right)
$$
\nd
This indicates that out of the Higgs fields given in
equations (\ref{v+1}--\ref{67}) the ones which give masses to
quarks and leptons, namely, ${V_{+1}}^i$ and ${V_{+4}}^t$
(which lie respectively inside $v$ and $\tilde v$ in the above
notation), do not couple to the background $\omega$ field on
${\mathbb{C}}P^2$. The leading term in their harmonic expansion
on ${\mathbb{C}}P^2$ will be a constant (independent of the
coordinates of ${\mathbb{C}}P^2$). Their masses will be tachyonic
and will be given by (\ref{mass}) for $\displaystyle\lambda=-
\frac{1}{2}$, i.e.
$\displaystyle M^2=-\frac{3}{2}\frac{1}{a_1^2}$.\\

The remaining fields ${V_{+1}}^t$ and ${V_{-i}}^4$ on the other
hand are located inside $u$ and they couple to the background
$\omega$-field. Their masses will receive contribution from
${\mathbb{C}}P^2$ and in principle can become non-tachyonic.
To verify this we need to evaluate the eigenvalues of
$\nabla^2_{{\mathbb{C}}P^2}$ on these fields. Their covariant
derivatives are
$$D{V_{+1}}^t =d{V_{+1}}^t+ \omega {V_{+1}}^t$$
$$D{V_{-4}}^i =d{V_{-4}}^i+ \omega {V_{-4}}^i $$
\nd
Since they couple with the same strength to the $\omega$-field
they will receive the same contribution from $\nabla^2_{{\mathbb{C}}
P^2}$. It turns out that the leading term in the expansion of any of
these fields on ${\mathbb{C}}P^2$ is a triplet of $SU(3)$ and $D^2$
acting on it is $\displaystyle -\frac{1}{a_2^2}$. Thus the total
mass$^2$ of such modes will be
$$
-\frac{3}{2}\frac{1}{a_2^2}+\frac{1}{a_2^2}=\frac{1}{a_1^2}
\left( -\frac{3}{2}+\frac{a_1^2}{a_2^2} \right)
$$
\nd
If $a_1$ and $a_2$ were independent we could choose $\displaystyle
(\frac{a_1}{a_2})^2
\geq \frac{3}{2}$ and make these fields non-tachyonic. If we insist
on the validity of the background Einstein equations then the ratio
of $\displaystyle\frac{a_1}{a_2}$ will be fixed. Equation
(\ref{ink}) leads to $\displaystyle (\frac{a_1}{a_2})^2=
\frac{12}{17}$.
\footnote{To obtain $\displaystyle (\frac{a_1}{a_2})^2=
\frac{12}{17}$  we need to use the following results, which
can be obtained by straightforward calculation,\\
$\displaystyle{\cal R}(S^2)=
\frac{1}{a_1^2}
1_{2\times 2}$,\
$\displaystyle{\cal R}({\mathbb{C}}P^2)=\frac{3}{2}\frac{1}{a_1^2}
1_{4\times 4}$,\
$\displaystyle\mbox{Tr}F^2_{S^2}=6 \frac{1}{a_1^4}$,\
and
$\displaystyle\mbox{Tr}F^2_{
{\mathbb{C}}P^2
}=\frac{51}{2} \frac{1}{a_2^2}$.}
With this value
unfortunately the above mass$^2$ is still negative. The vacuum
expectation value of these fields will break the color $SU(3)$.\\

One way to change the ratio $\displaystyle\frac{a_1}{a_2}$ is to
couple a $U(1)$ gauge field to gravity in $D=10$. This $U(1)$ will
{\it not} couple to anything else. In particular the fermions will
be neutral under it, so the spectrum of the chiral fermions will be
unaltered. Its sole effect will be to add an extra term to the
right hand side of Einstein equations. In particular (\ref{ink})
will be replaced by
$$
{\cal R}_{{\hat m}{\hat n}}=\frac{\kappa^2}{g^2}\mbox{Tr}
F_{{\hat m}{\hat r}}{F_{\hat n}}^{\hat r}
+\frac{\kappa^2}{{g'}^2}\mbox{Tr}
{F'}_{{\hat m}{\hat r}}{{F'}_{\hat n}}^{\hat r}
$$
\nd
where $F'$ and $g'$ refer to the extra $U(1)$ system. Now if we set
$$
A'=\frac{n'}{2}(\mbox{cos}\theta
-1)d\varphi + q'i\omega
$$
\nd
where
$n'$ and $q'$ are real numbers, the ratio of $a_1/a_2$
will turns out to be
$$
\displaystyle
\frac{a_1^2}{a_2^2}=
\frac{36+3{n'}^2\frac{g^2}{{g'}^2}}{51+2{q'}^2\frac{g^2}{{g'}^2} }
$$
\nd
There is a big range of parameters for which  $a_1/a_2\geq 3/2$ .


\section{Other Scalars \label{section7}}


The components of the gauge field fluctuations tangent to
${\mathbb{C}}P^2$ will also give rise to infinite tower of
Kaluza--Klein modes which will be scalars fields in $D=4$. These
modes will belong to unitary representations of $SU(2)\times
SU(3)$. If there is a tachyon Higgs among them they will break
$SU(3)$. We need to verify that this does not happen. To this end
we denote these fields by $V_a$, where $a$ is tangent to
${\mathbb{C}}P^2$, and write those terms in the bilinear part of
$\mbox{Tr}F_{MN}F^{MN}$ which contains $V_a$. In this section we
are considering only the $U(6)$ model. The $V_a$ are $5\times 5$
Hermitian matrices. After some manipulation and the imposition if
the $D=10$ background gauge condition $D_MV^M=0$, the bilinear
terms of interest to us can be written as
\begin{equation}
S_2= -\frac{1}{2g^2}\int d^{10}x \mbox{Tr}\{
2V_a(-\partial^2-D_m^2-D_{\hat m}^2+\frac{3}{2}\frac{1}{a_2^2})
V^a
+4iV^a[{\bar F}_{ab},V^b] \}
\label{s2}
\end{equation}
\nd
where $D_m$ and $D_{\hat m}$ are respectively the covariant
derivatives on $S^2$ and ${\mathbb{C}}P^2$ and
\begin{equation}
D_m V_a=\partial_m V_a-\frac{i}{2} \omega_m [n,V_a]
\end{equation}
\begin{equation}
D_{\hat m} V_a=\nabla_{\hat m}V_a-\frac{i}{2} \omega_{\hat m}[q,V_a]
\label{ru3}
\end{equation}
\nd
$\nabla_{\hat m}$ is the Riemann covariant derivative on
${\mathbb{C}}P^2$. The contribution of $D_m^2$ on each $SU(2)$
mode of $V_a$ will simply be $\displaystyle
\frac{1}{a_1^2}[l(l+1)-\lambda^2]$,
$l\geq |\lambda|$ where $\lambda$ represents the isohelicities of
various components of $V_a$, $\lambda({V_{ai}}^j)=\lambda (
{V_{+i}}^j)-1$, where $\lambda ({V_{+i}}^j)$ are given
in equation (\ref{iso}).\\

To work out the contributions of $D_{\hat m}^2$ and the commutator
term $[{\bar F}_{ab},V^b]$, we represent $V_a$ as in (\ref{V}),
i.e.
\begin{equation}
V_a=
\left(\begin{array}{c|c}
\;\;v_a\;\;&\;\;u_a\;\;\\
\hline
\;\;{\tilde u_a}\;\;& \;\;{\tilde v_a}\;\;\end{array}
\right)
\label{Va}
\end{equation}
\nd
where $v_a$, $\tilde v_a$,  $u_a$, $\tilde u_a$
each is a $3\times 3$ matrix.
Then
\begin{equation}
[q,V_a]=
\left(\begin{array}{c|c}
\;\;0\;\;&\;\;u_a\;\;\\
\hline
\;-{\tilde u_a}\;\;&\;\; 0\;\;\end{array}
\right)
\end{equation}
\nd
This indicates that the commutator terms in (\ref{s2}) and
(\ref{ru3}) do not contribute to $D_{\hat m}v_a$ and
$D_{\hat m}{\tilde v}_a$. Thus $D_{\hat m}$ acting on these fields
is just the Riemannian Laplacian acting on vectors and its
contribution to the masses of these fields will be non-tachyonic.\\

The only fields we need to be concerned about are those in
$u_a$. To  analyze the contribution of these terms we
introduce $2$ complex
$SU(2)$ vectors $u_\alpha$ and $u'_\alpha$ defined by
\begin{equation}
\left\{\begin{array}{l}
u_{\underline{1}}=\frac{1}{\sqrt{2}}(u_6+iu_7) \\
u_{\underline{2}}=\frac{1}{\sqrt{2}}(u_8+iu_9) \\
\end{array} \right.
\;\;\;\;\;\;\;\;\;\;\;\;\;\;
\left\{\begin{array}{l}
u'_{\underline{1}}=\frac{1}{\sqrt{2}}(u_6-iu_7) \\
u'_{\underline{2}}=\frac{1}{\sqrt{2}}(u_8-iu_9) \\
\end{array} \right.
\end{equation}
\nd
where $6$, $7$, $8$, and $9$ are directions tangent to
${\mathbb{C}}P^2$. In terms of these new fields the $u_a$ part
of (\ref{ru3}) can be rewritten as
\begin{equation}
D_{\hat m} u_\alpha=
(\partial_{\hat m}+i\omega_{\hat m}^i\frac{\sigma^i}{2}
-i\frac{5}{2} \omega_{\hat m}) u_\alpha
\label{x4}
\end{equation}
\begin{equation}
D_{\hat m} u'_\alpha=
(\partial_{\hat m}+i\omega_{\hat m}^i\frac{\sigma^i}{2}
-i\frac{5}{2} \omega_{\hat m}) u'_\alpha
\label{x5}
\end{equation}
\nd
The contribution of $D^2_{\hat m}$ on $u_\alpha$ and $u'_\alpha$
will again be positive. \\

Finally we need to evaluate the contribution of $2i\mbox{Tr}V^a
[{\bar F}_{ab}, V^b]$ to the masses of $u_\alpha$ and $u'_\alpha$.
After some calculation this turns out to be
\begin{equation}
2i\mbox{Tr}V^a
[{\bar F}_{ab}, V^b]=
\frac{2}{a^2_2}\mbox{Tr}(u^\dag_
\alpha u_\alpha-{u'}^\dag_\alpha {u'}_\alpha)
\label{fi}
\end{equation}
\nd
It is seen that the contribution of this term to the $u_\alpha$
mass is non-tachyonic. However, it makes a negative contribution
to the mass$^2$ of $u'_\alpha$ field. Upon substitution of the
above in (\ref{ru3}) we find out that the negative contribution in
(\ref{fi}) is off-set by the $\displaystyle
\frac{3}{2}\frac{1}{a_2^2}$ term in
equation (\ref{s2}), with the result that $u'_\alpha$ is also
non-tachyonic. \\

We thus conclude that all the tachyonic Higgs are singlets of
$SU(3)$ and doublets of $SU(2)$.


\section{Massless Scalars and Loop induced Hierarchy
\label{massless}}


So far we have been discussing tachyonic mass of the scalar
particles at the tree level of the effective $4$ dimensional
theory. The natural scale of this mass and therefore also of the
symmetry breaking is the compactification scale. This is few
order of magnitude above the electroweak symmetry breaking scale
of a 200 hundred GeV. It will be very desirable if we could find
a mechanism to lower the scale of the tachyonic mass. An obvious
idea is if the tree level mass of the scalars is zero and they
obtain their tachyonic value as a consequence of loop effects.
Our theory is of course a non renormalizable one, at least in
conventional sense. However, the Higgs mass is
controlled by
$1/a$ due to higher dimensional gauge invariance.
Our main point is that the sign of the one loop induced effective
mass will depend on the imbalance between the contribution of
fermions and bosons. By a judicious choice of the fermionic
degrees of freedom this sign can be made tachyonic. Any way
whatever the justification the first step in implementing this
idea is to find tree level massless scalars in the spectrum of the
effective four dimensional theory. Unlike the massless chiral
fermions whose presence is dictated by the topology of the gauge
field in compact subspace, to verify the existence of the massless
scalars in the spectrum requires more detailed analysis of the
mass spectrum and should be carried out separately for each case.
In this section we give an example of a model in $D=10$ in which a
monopole background on the $S^2 \times S^{'2}\times S^{''2}$
internal space leads to  massless scalars transforming non
trivially under the $SU(2)\times SU(2)\times SU(2)$ isometry group
of the internal space. This example which was  is only for
illustrative purpose
and is not going to be used for a realistic model building.\\

We start from a $U(N)$ gauge theory in $10$ dimensions and
consider a solution of equations (\ref{ink}) in which the internal space
is $S^2\times S^{'2}\times S^{"2}$. In the notation of previous
section we denote the magnetic charge matrices on the three
$S^{2}$'s by  $n$  $n'$ and $n"$. Denoting all the quantities on
$S{'2}$ with a prime our ansatz for the gauge field becomes
$$
A = {n\over 2}(cos \theta -1)d\phi + {n'\over 2}(cos \theta'
-1)d\phi' +{n''\over 2}(cos \theta'' -1)d\phi''
$$

The structure of the charge matrices will determine the unbroken
subgroup of $U(N)$. As before we shall take them to be $N\times
N$ diagonal real matrices.\\

The scalars of interest for us are those components of the
fluctuations of the vector potential which are tangent to
$S^2\times S^{'2}\times S^{"2}$ and are in the directions of
perpendicular to the Cartan subalgebra of $U(N)$. Consider the
field $V_{-i}^{j}$ tangent to $S^2$.\\

The masses of these fields can be calculated using the
appropriate modification of equation (\ref{s2}). The result is\\

$
\displaystyle
S_2= -\frac{1}{2g^2}\int d^{10}x  \Big\{
(V_{-i}^{j})^*(-\partial^2-D^2-D^{'2}- D^{''2}+{1\over a^2})
V_{-i}^{j}$
\begin{equation}
\;\;\;\;\;
\;\;\;\;\;\;
\;\;\;\;\;
\;\;\;\;\;\;\;\;\;\;\;
\;\;\;\;\;\;
\;\;\;\;\;
\;\;\;\;\;
\;\;\;\;\;
\;\;\;\;\;
\;\;\;\;\;
-{1\over a^2}(V_{-i}^{j})^*(n_{i}- n_{j})V_{-i}^{j}
\Big\}
\label{s22}
\end{equation}
\nd
where $D^2$, $D^{'2}$ and $D^{''2}$ are the appropriate
Laplacian on the three $S^2$'s. The eigenvalues of these
Laplacians are basically determined from the isohelicities of
$V_{-i}^{j}$  which are given by\\

$
\lambda(V_{-i}^{j}) = -1 +{1\over 2}(n_i-n_j),\;\;\;\;\;
\lambda^{'}(V_{-i}^{j}) = {1\over 2}(n_i^{'}-n_j^{'}),\;\;\;\;\;
\lambda^{''}(V_{-i}^{j}) ={1\over 2}(n_i^{''}-n_j^{''})$\\

Similar expressions can be written for the bilinear parts of the
fields tangent to $S^{'2}$ and $S^{''2}$.\\

For our illustrative example we consider an $n$ matrix which has
only the elements $n_1$ and $n_2$ different from zero and
such that $n_1 -n_2 \geq2$. Then $\lambda (V_{-1}^{2})\geq 0$ and
according to our general rule the leading mode in this field can
be tachyonic.  The question we would like to answer is if by an
appropriate choice of magnetic charges we can make the mass of
this field to vanish. It is not difficult to write down the
formula for the masses of the infinite tower of modes of
$V_1^2$. These are given by
$$a^{2}M^2= l(l+1) - \lambda^2 + {a^2\over a^{'2}}
(l'(l'+1) -\lambda^{'2})+ {a^2\over a^{''2}}
(l''(l''+1) -\lambda^{''2}) + 1-(n_1-n_2)$$

To verify the existence of a massless mode
first we employ the background
equations (\ref{ink}) to obtain the ratios
$$
{a^2\over a^{'2}}= {{\mbox{Tr}n^2}\over {\mbox{Tr} n^{'2}}}
,\;\;\;\mbox{and}\;\;\;
{a^2\over a^{''2}}= {{\mbox{Tr} n^2}\over
{\mbox{Tr} n^{''2}}}.$$
\nd
It is seen that for the choice of $n'_1 - n'_2 =
n_1- n_2$ , $\mbox{Tr} n^2= \mbox{Tr} n^{'2}$
and $ n_{1}^{''}- n_{2}^{''}=0 $
the leading mode is  indeed massless. For this choice there will
of course be a similar massless mode in the fluctuations $V_{-
1}^{'2}$ tangent to $S^{'2}$. The $SU(2)\times SU(2)\times SU(2)$
quantum numbers of these modes will be \\
$(l={1\over 2}(n_1 -n_2)-1,
l'={1\over 2}(n_1 -n_2), 0)$, and
$l=({1\over 2}(n_1 -n_2),
l'={1\over 2}(n_1 -n_2)-1, 0)$,
respectively.  We can make all
other modes to have positive masses by appropriate choices of the
remaining magnetic charges.


\section{Summary and outlook \label{summary}}


In this paper we argued that the Higgs scalars  of the
$4$-dimensional spontaneously broken gauge theories have their
origin in the extra components of a Yang-Mills potential in
$4+N$ dimensions. For this idea to be useful and tenable
it should be shown that there is a mechanism through
which these scalars break the gauge symmetries
spontaneously. We showed that their coupling to a
background magnetic monopole field is one such mechanism.
This coupling gives a tree level tachyonic mass to these
scalars and also makes it possible that their leading
mode in the harmonics of $S^2$ to belong to the doublet
representation of the isometry group of $S^2$, which we
identified with $SU(2)_L$ of the electroweak theory. The
presence of left handed fermionic  doublets and right handed
fermionic singlets justifies this identification.  We gave a
simple rule to identify the Kaluza--Klein modes which could
trigger spontaneous symmetry breaking in the effective
4-dimensional theory.  We constructed two examples with
gauge groups
$SU(3)$ and
$U(6)\times U(1)$ in $D=6$ and
$D=10$, respectively.  The first example leads to an effective
left - right symmetric type model in $D=4$ with the gauge group
$SU(2)_L\times SU(2)_R\times U(1)$. We worked out the full $D=4$
effective action for this model.\\

In the second example the internal space is $S^2\times CP^2$ with
a magnetic monopole configuration on $S^2$ and a $U(1)$ instanton
on ${\mathbb{C}}P^2$.  The modes which we retained in the effective $D=4$
theory are chiral fermions and tachyonic scalars to be identified
as the Higgs fields. It is straightforward to work out the
effective $D=4$ action for this example as well.\\

Although in the present paper our intention was not to recover
the standard model of particle physics from a higher dimensional
theory, the chiral spectrum of leptons and quarks and the
representation content of the effective Higgs fields as well as
their Yukawa couplings and potential in our illustrative example
are reasonably realistic to warrant further study of examples
like the ones of this paper. For instance we need to understand
how the $U(1)$ symmetries in the $U(6)$ model are going to be
broken. The vacuum expectation value of the Higgs in the lepton
and the quark sectors will break only two out of the four
effective $D=4$ $U(1)$ symmetries. Presumably some loop effects
will generate condensates of some composite objects which are
charged under these $U(1)$'s and thereby break these symmetries
dynamically.\footnote{ The $SU(2)\times SU(2)$ subgroup of $U(6)$
will of course be broken by our Higgs fields.}\\

It is worth noting that since $\pi_{10}(U(6)\times U(1))$ is
trivial there will be no global Witten anomalies. The
perturbative anomalies on the other hand can be removed by the
Green Schwarz type mechanism explained at the end of section
\ref{anomaly1}.\\

For obvious phenomenological reasons one would rather prefer to
have a (mild) hierarchy of scales $M_W < 1/a_1$. This might be
possible if the tachyonic mass could be induced as a result of a
one-loop effect. A necessary precondition for this is the
presence of massless scalars in the tree level spectrum of the
effective $4$ dimensional theory. To show that this is indeed
possible we considered a $10$ dimensional $U(N)$ model with a
solution which posits  magnetic monopoles with charge matrices
$n$ , $n'$ and $n''$ on each one of the three $S^2$'s comprising
the 6-dimensional compact internal space. For a judicious choice
of some of the elements of these matrices we do indeed obtain
massless scalars in non trivial representations of the unbroken
gauge group $SU(2)\times SU(2)\times SU(2)\times G$, where $G$ is
the subgroup of $U(N)$ which is left unbroken by our background
solution. As a general rule the loop effects are expected to
induce a non zero mass for these scalars and it is possible to
verify in detail that an appropriate choice of the fermionic
couplings the induced mass will indeed be tachyonic. The quartic
self coupling of these scalars can be worked out as we did for the
$SU(3)$ model of section \ref{examples}. \footnote{ After we had
submitted our paper to the arXiv, a paper,
\cite{st}, appeared in
which the hierarchy problem in the context of
higher dimensional theories is addressed
in a different way.}\\

\nd
{\bf Acknowledgments}

\vspace{0.1in}

We are grateful to Nima Arkani-Hamed,
Goran Senjanovi\'c, Alexei Smirnov and
George Thompson for discussion. GD thanks High Energy Group
of ICTP for hospitality. The work of GD was supported in
part by David and Lucille Packard Foundation Fellowship for
Science and Engineering, by Alfred P. Sloan foundation
fellowship and by NSF grant PHY-0070787.

\end{document}